\documentstyle{aipproc}

\input epsf.tex

\def\APJ#1#2#3{Ap. J. {\bf #1}, #2 (#3)}

\newcommand{\postscript}[2]{\setlength{\epsfxsize}{#2\hsize}
   \centerline{\epsfbox{#1}}}

\newcommand{\tb}{\tan\beta}

\newcommand{\gev}{\text{GeV}}
\newcommand{\tev}{\text{TeV}}
\newcommand{\cm}{\text{cm}}
\newcommand{\km}{\text{km}}
\newcommand{\s}{\text{s}}
\newcommand{\yr}{\text{yr}}

\newcommand{\Omegachi}{\Omega_{\chi}}
\newcommand{\be}{\begin{equation}}
\newcommand{\ee}{\end{equation}}
\newcommand{\etal}{{\em et al.}}

\newcommand{\rem}[1]{#1} 

\begin{document}
\title{Dark Matter Implications \\
for Linear Colliders
\rem{\thanks{Talk presented at Linear Collider Workshop 2000, 
24-28 October 2000, Fermilab, Batavia, Illinois, USA.}}
}

\author{Jonathan L.~Feng
\rem{\thanks{Electronic address: jlf@mit.edu}}
\thanks{This work was supported in part by funds provided by the
U.~S.~Department of Energy under cooperative research agreement
DF--FC02--94ER40818.}
}

\address{Center for Theoretical Physics, Massachusetts Institute
of Technology\\ Cambridge, MA 02139 USA}

\maketitle

\rem{\vskip-.2in}
\begin{abstract}
The existence of dark matter is currently one of the strongest
motivations for physics beyond the standard model. Its implications
for future colliders are discussed. In the case of neutralino dark
matter, cosmological bounds do not provide useful upper limits on
superpartner masses.  However, in simple models, cosmological
considerations do imply that for supersymmetry to be observable at a
500 GeV linear collider, some signature of supersymmetry must appear
{\em before} the LHC.
\end{abstract}

\rem{\vskip-3.9in
\noindent
MIT--CTP--3065 \hfill hep-ph/0012277
\vskip3.5in
}
\section*{Introduction}

In evaluating any large-scale future project in high energy physics,
the critical question at present is its ability to discover and
explore physics beyond the standard model.  While many theoretical
motivations for such physics exist, one of the most compelling
phenomenologically (along with neutrino oscillations) is the evidence
for dark matter.  The energy density of luminous matter in the
universe is $\Omega_{\text{lum}} \approx 0.005$.  At the same time,
measurements of mass in galactic clusters, expected to be the largest
virialized structures in the universe, require $\Omega_m \approx
0.2$~\cite{Carlberg}, and recent measurements of supernovae
luminosities and CMB anisotropies imply $0.2 \lesssim \Omega_m
\lesssim 0.4$~\cite{Jaffe}.  Additional observations require most of
the dark matter to be cold and non-baryonic.  No particles of the
standard model (even suitably extended to include neutrino masses) are
even remotely plausible candidates.

Among the most promising particles beyond the standard model are two
motivated by fine-tuning problems --- the lightest neutralino and the
axion.  Dark matter may be composed of either, neither, or both.
Axion dark matter is of little relevance for high energy colliders
(but is the subject of another vigorous experimental
program~\cite{Rosenberg:2000wb}).  In contrast, neutralino dark matter
has strong implications for colliders and has even been argued to
provide stringent upper bounds on superpartner masses.  We will see
that this is overly optimistic even in simple
models~\cite{Feng:2000gh}.  However, we will find that the requirement
of neutralino dark matter does strongly constrain parameter space.  In
simple models like minimal supergravity, if superpartners are
accessible at a 500 GeV linear collider, some hint of supersymmetry
must appear before the LHC~\cite{Feng:2000zu}.

\section*{Neutralino Dark Matter}

The lightest neutralino $\chi$, is well-known to be an excellent dark
matter candidate in $R$-parity conserving supergravity
models~\cite{Goldberg:1983nd}.  In addition to being stable, neutral,
and non-baryonic, its annihilation cross section gives, very roughly,
the desired thermal relic density:
\begin{equation}
\Omegachi \approx \frac{10^{-10}~\gev^{-2}} {\langle \sigma_A v
\rangle} \sim \frac{10^{-10}~\gev^{-2}} {(\alpha^2/ m_W^2) \times 0.1}
\sim 0.1 \ .
\end{equation}

To go beyond such general statements, one must specify precisely all
of the many supersymmetry parameters that determine neutralino
properties.  The full parameter space is unwieldy.  However, many
important insights may be gained by considering the simple example of
minimal supergravity.  In this framework, the weak scale values of the
Bino and Wino masses satisfy $2M_1 \approx M_2$, and $|\mu|$ is fixed
by electroweak symmetry breaking and the universal scalar mass
$m_0$~\cite{Feng:2000mn}:
\begin{equation}
\frac{1}{2} m_Z^2 \approx -0.04\; m_0^2 + 8.8\; M_1^2 - \mu^2\ .
\label{mz2}
\end{equation}

\begin{figure}[tbh]
\postscript{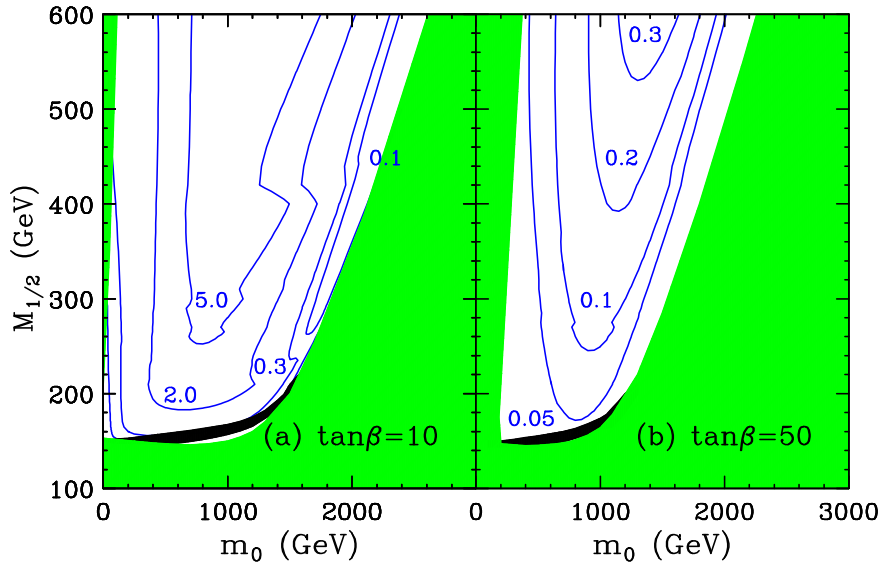}{0.78}
\caption
{Contours of $\Omegachi h^2$~\protect\cite{Feng:2000gh}.  We fix
$A_0=0$ and $\mu>0$, and choose representative values of $\tb$ as
indicated.  The shaded regions are excluded by the requirements of a
neutral LSP (left) and the 103 GeV chargino mass bound (right and
bottom). In the black region, neutralinos annihilate through the light
Higgs pole. (Heavy Higgs poles also play a role in limited regions
with $\tb=50$ and $m_0 < 1~\tev$.)  Effects of co-annihilation,
important along the boundaries of the excluded regions, have not been
included.  }
\label{fig:omegaHsq}
\end{figure}

The neutralino thermal relic density in minimal supergravity is given
in Fig.~\ref{fig:omegaHsq}.  In the region $m_0 \lesssim 1~\tev$, an
upper bound of, say, $\Omegachi h^2 \lesssim 0.3$ leads to upper
bounds on both the universal scalar and gaugino masses.  This is easy
to understand qualitatively: in this region, Eq.~(\ref{mz2}) implies
$M_1 < \mu$, and so $\chi \approx \tilde{B}$.  Bino-like dark matter
annihilates dominantly to fermion pairs through $t$-channel
$\tilde{f}$ exchange, and so $\Omegachi h^2$ grows as $m_0$ increases.
This line of reasoning seemingly leads to the conclusion that
cosmology implies that some superpartners must be within reach of a
TeV linear collider.

However, for $m_0 \gtrsim 1~\tev$, it is possible to satisfy
Eq.~(\ref{mz2}) with $M_1 \sim \mu$, and $\chi$ may be a
gaugino-Higgsino mixture.  Such neutralinos dominantly annihilate
through $t$-channel charginos and neutralinos to gauge bosons, and so
another branch of parameter space with cosmologically-preferred
$\Omegachi h^2$ exists.  This branch extends to $m_{\chi} \sim
2.5~\tev$~\cite{Feng:2000gh}, where unitarity ultimately limits
$\Omegachi h^2$.  Thus, while cosmology does provide upper bounds on
superpartner masses, the upper bounds are not stringent enough to
guarantee supersymmetry at any foreseeable linear collider.

The $m_0 \gtrsim 1~\tev$ branch, often neglected, is comparable in
size to the conventional $m_0 \lesssim 1~\tev$ branch and has
significant virtues~\cite{Feng:2000bp}: undesirable contributions to
proton decay and electric dipole moments are suppressed, and heavy top
and bottom squarks naturally predict Higgs boson masses of $115~\gev
\lesssim m_h \lesssim 120~\gev$, the range preferred by current
data. In addition, in a sense precisely defined in
Ref.~\cite{Feng:2000mn}, no additional fine-tuning is required as a
result of the interesting `coincidence' that the top quark mass is
$m_t \approx 180~\gev$~\cite{Feng:2000mn,Feng:2000hg}.

\section*{Prospects for Supersymmetry Discovery}

\begin{figure}[tbh]
\postscript{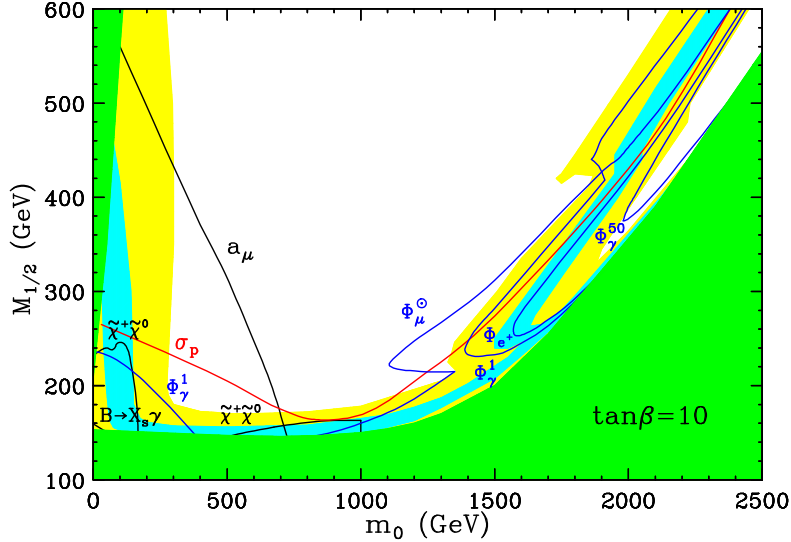}{0.7}
\caption
{Reaches of various astrophysical and particle physics experiments
expected by 2006~\protect\cite{Feng:2000zu}.  The excluded regions are
as in Fig.~\protect\ref{fig:omegaHsq}.  In the light (dark) shaded
region, $0.025 \protect\lesssim \Omegachi h^2 \protect\lesssim 1$
($0.1 \protect\lesssim \Omegachi h^2 \protect\lesssim 0.3$). The
regions probed extend the curves toward the excluded regions.  }
\label{fig:reach10}
\end{figure}

If neutralinos account for a significant fraction of the dark matter,
many experiments have the potential to discover supersymmetry.  On the
$m_0 \lesssim 1~\tev$ branch, traditional particle physics experiments
are sensitive.  On the other hand, for large $m_0$, many dark matter
searches are especially powerful. The projected reaches of both
particle physics experiments and dark matter searches by the year 2006
are given in Fig.~\ref{fig:reach10}.  The observable signals,
associated experiments, and expected sensitivities are given in
Table~\ref{table:comp}.

\begin{table}[tbh]
\caption{Supersymmetric signals and experimental sensitivities assumed
in Fig.~\ref{fig:reach10}.
\label{table:comp}
}
\begin{tabular}{llll}
 Observable 
  & Type
   & Sensitivity
    & Experiment(s)   \\  \hline
 $\tilde{\chi}^{\pm} \tilde{\chi}^0$
  & Collider
   & See Ref.~\cite{Feng:2000zu}
    & Tevatron: CDF, D0  \\ 
 $B \to X_s \gamma$
  & Low energy
   & $|\Delta B(B\rightarrow X_s\gamma)| < 1.2\times 10^{-4}$
    & BaBar, BELLE     \\
 Muon MDM
  & Low energy
   & $|a_{\mu}^{\text{SUSY}}| < 8 \times 10^{-10}$
    & Brookhaven E821  \\
 $\sigma_{\text{proton}}$
  & Direct DM
   & $\sim 10^{-8}$ pb (See Ref.~\cite{Feng:2000zu})
    & CDMS, CRESST, GENIUS \\
 $\nu$ from Earth
  & Indirect DM
   & $\Phi_{\mu}^{\oplus} < 100~\km^{-2}~\yr^{-1}$
    & Amanda, Nestor, Antares \\
 $\nu$ from Sun
  & Indirect DM
   & $\Phi_{\mu}^{\odot} < 100~\km^{-2}~\yr^{-1}$
    & Amanda, Nestor, Antares \\
 $\gamma$ (gal. center)
  & Indirect DM
   & $\Phi_{\gamma}(1) < 1.5\times 10^{-10}~\cm^{-2}~\s^{-1}$
    & GLAST \\
 $\gamma$ (gal. center)
  & Indirect DM
   & $\Phi_{\gamma}(50) < 7\times 10^{-12}~\cm^{-2}~\s^{-1}$
    & MAGIC \\
 $e^+$ cosmic rays
  & Indirect DM
   & $(S/B)_{\text{max}} < 0.01$
    & AMS-02           
\end{tabular}
\end{table}

In addition to the complementarity of the particle and astrophysical
experiments, it is notable that all of the cosmologically-preferred
parameter space accessible to a 500 GeV linear collider will lead to
at least one hint of supersymmetry before the LHC begins operation.
This conclusion applies for all $\tb$ in minimal supergravity, and its
qualitative structure suggests that similar conclusions will remain
valid in alternative frameworks.

\vspace*{.15in}

I thank K.~Matchev and F.~Wilczek for collaboration in the work
described here.

\end{document}